\documentclass[twocolumn, twoside,prl]{revtex4}

\usepackage{amsmath}
\usepackage{amssymb}
\usepackage{latexsym}
\usepackage{revsymb}
\usepackage{epsfig}

\newcommand{\barr}{\begin{eqnarray}}
\newcommand{\earr}{\end{eqnarray}}
\newcommand{\ra}{\rangle}

\newcommand{\beq}{\begin{equation}}
\newcommand{\eeq}{\end{equation}}

\begin{document}

\title{Maximally efficient quantum thermal machines: The basic principles}

\author{Sandu Popescu} \affiliation{H. H. Wills Physics Laboratory, University of Bristol$\text{,}$ Tyndall Avenue, Bristol, BS8 1TL, United Kingdom }
\
\date{\today}
\begin{abstract}
Following the result by Skrzypczyk et al., arXiv:1009.0865, that certain self-contained quantum thermal machines can reach Carnot efficiency, we discuss the functioning of self-contained quantum thermal machines and show, in a very general case, that they can reach the Carnot efficiency limit. Most importantly, the full analytical solution for the functioning of the machines is not required; the efficiency can be deduced from a very small number of fundamental and highly intuitive equations which capture the core of the problem.
\end{abstract}

\maketitle

In a very recent work \cite{lps} two fundamental questions were raised about thermal machines. The first question was whether or not there exists a fundamental limitation to the size of (quantum) thermal machines (where size is measured in the number of quantum states the machine). The second question refers to the efficiency of the machines: is there any complementarity between size and efficiency? That is, can the Carnot efficiency be reached by machines with only very few quantum states? The first question was answered in \cite{lps} where the smallest refrigerator was designed: there is effectively no lower limit on the number of states. The second question was answered in \cite{carnot} where it was shown that there is no trade-off between size and efficiency and that the smallest possible refrigerator can reach the Carnot limit.

The results in \cite{carnot} however are based on rather complicated computations, involving solving for the exact analytical solution. All these computation mysteriously simplify in the end. Here we revisit the problem and
show that finding the entire analytical solution (which depends on all the parameters of the problem and on the details of the interaction with the environment) is not necessary. Instead we
formulate main principles that govern the functioning of our quantum thermal machines; these principles capture the core of the problem and lead to the Carnot efficiency in a clear, straightforward and very intuitive manner.

Obviously, the above results do not come into a vacuum. During the last couple of years there has been a lot of interest in the functioning of quantum thermal machines \cite{I1}-\cite{Bard01}, with major results being obtained. Here however we are specifically interested in accounting for {\it all} the degrees of freedom of the machine, for all its states. Hence we are considering fully self-contained machines and we do not allow, explicitly or implicitly, for any external source of work. In particular, we do not allow for time dependent Hamiltonians or prescribed unitary transformations. All that our machines are allowed is access to heat baths.

\section{A quantum self-contained refrigerator}

\bigskip
\noindent
{\bf The Model} The refrigerator we consider consists of three qubits, 1,2 and 3, each in contact with a different heat bath $T_1>T_2>T_3$. Qubit 1 is in contact with the "hot" bath $T_1$, qubit 2 is in contact with a "room temperature" bath, $T_2$, and qubit 3 is in contact with the "cold" bath $T_3$. The refrigerator works by taking heat from the cold bath and dumping it at the room temperature. Free energy to run the fridge is provided by the hot bath in conjunction with the room temperature one: energy flows from the hot bath into the room temperature one.

The defining characteristics of the refrigerator are:
\begin{itemize}
\item The free hamiltonians of each qubit are chosen such that the energy differences $E_i$ between the excited state $|1\ra_i$ and ground state $|0\ra_i$ of each qubit are chosen such that
\beq E_1+E_3=E_2.\label{energies}\eeq
Hence the states $|010\ra$ and $|101\ra$ are degenerate. (Here we used the simplified notation $|0\ra_1|1\ra_2|0\ra_3=|010\ra$ and so on.)
\item The interaction Hamiltonian is weak relative to the free Hamiltonian. Apart from this, the interaction hamiltonian is completely arbitrary.
\end{itemize}
Given that the interaction Hamiltonian is weak relative to the free one, we can take with a very good approximation the free Hamiltonian to define the total energy of each qubit. Furthermore,since the interaction Hamiltonian is weak, all it can do is to produce transitions between the two degenerate energy levels $|010\ra$ and $|101\ra$. Even if the interaction Hamiltonian has terms that couple different states, in the weak coupling regime all other transitions are suppressed.

We note however, that even though the interaction Hamiltonian is weak, this doesn't mean that the state of the system is necessarily approximately equal to the state in the absence of interaction. In fact the state can be significantly different; it all depends on the exact parameters describing the interaction of the qubits with the heat baths. A particular example is explicitly calculated in \cite{lps}.

\bigskip
\noindent
{\bf Working regime} The whole idea of the refrigerator is to try and cool spin 3 to a temperature lower than that of the cold bath with which it is in contact. When this is done, it will draw heat from the cold bath.

Cooling spin 3 is tantamount to increasing the probability to find it in the ground state. Hence all we have to do is to try and enhance the probability of the transition $|101\ra\rightarrow|010\ra$ (in which qubit 3 goes from the excited state to the ground) over the probability of the reverse transition $|010\ra\rightarrow|101\ra$. This is done by arranging that the probability to find the system in the state $|101\ra$ in the absence of interaction is larger than the probability to find the system in state $|010\ra$, i.e. when

\beq e^{-{{E_1}\over{kT_1}}} e^{-{{E_3}\over{kT_3}}}> e^{-{{E_2}\over{kT_2}}}\label{working_regime_fridge}\eeq
which leads to
\beq{{E_1}\over{T_1}}+{{E_3}\over{T_3}}<{{E_2}\over{T_2}}\label{working_regime2_fridge}.\eeq

Equation (\ref{working_regime2_fridge}) defines thus the working regime.

We note that equations (\ref{working_regime_fridge}) and (\ref{working_regime2_fridge}) refer to the populations of the different levels in the {\it absence} of interaction. When the interaction is turned on, the populations change. In particular, the actual temperatures of the three qubits will be different from the ones of their surroundings. One may therefore wander why it is the population in the absence of interaction that defines the working regime and not what happens when the interaction is on. The reason is that if the initial populations are as stated, when the interaction is turned on, the refrigerator starts cooling. Qubit 3 will become colder than the cold bath, so draws heat from it, qubit 2 becomes warmer than room temperature (since it is pushed into the excited state) so it will dump heat into the room and qubit 1 will become cooler than the hot bath, so it extracts heat from it, to keep the refrigerator working. Due to the interaction qubits 3 and 1 will continue to cool while qubit 2 will continue to warm-up until, after a transient period, they will each stabilize at a working temperature which depend on the parameters of the problem. Whatever the details of these final temperatures are, they are however in the right order relative to their environments so that the system works as a refrigerator.

\bigskip
\noindent
{\bf Heat flow} The qubits exchange energy with their environment as well as with each other. What interests us here is the energy that is exchanged between them - at equilibrium the total energy of each qubit is constant, so the energy a qubit extracts from the environment is equal to the energy passed to the other qubits. Due to the fact that the only thing the interaction does is to make transitions between the states $|101\ra$ and $|010\ra$, the energy gains and loses of the qubits due to interaction are constraint:  Whenever qubit 1 losses energy $E_1$, qubit 3 must lose energy $E_3$ and qubit 2 must gain energy $E_2$ and vice-versa, when qubit 1 gains $E_1$, qubit 3 gains $E_3$ and qubit 2 loses $E_2$. How often a forward or a backward transition between $|101\ra$ and $|010\ra$ occurs again depends on the parameters of the problem: the exact interaction Hamiltonian, the strength of coupling of each qubit with its thermal bath, the exact model of the interaction with the bath. Hence, without these details we cannot tell what the heat exchange rates are. That is, we cannot tell how much heat is extracted or dumped into the baths {\it per unit time}.

However, and this is the key element of the entire argument, due to the constraint on the energy exchanges between the qubits, the  ratio of the heat exchanges with the baths is independent from all the details and it is simply the same as the ratio of the energy exchanges between the qubits, which is the ratio between the energy levels:

\beq Q_1 : Q_2 : Q_3 = E_1 : E_2 : E_3\label{ratios}\eeq

\bigskip
\noindent
{\bf Reversibility} The main question raised in this paper is that of the maximum efficiency of the refrigerator. As in all thermal machines, maximal efficiency is obtained at reversibility. For this we must ensure that the forward transition, that results in cooling qubit 3, is infinitesimally close to the reverse transition, that warms qubit 3. Hence the reversibility condition means that the $>$ sign in equations (\ref{working_regime_fridge}) and (\ref{working_regime2_fridge}) has to be replaced by equality. That is, at reversibility

\beq{{E_1}\over{T_1}}+{{E_3}\over{T_3}}={{E_2}\over{T_2}}\label{working_regime3_fridge}.\eeq

Incidentally we note that just running the refrigerator more slowly by simply making the interaction Hamiltonian weaker is not enough. This will not ensure reversibility. The actual design of the fridge (i.e. the energy levels) has to be matched to the working temperatures, as in (\ref{working_regime3_fridge}). This is similar to the case of macroscopic devices. Indeed, consider a refrigerator consisting of a cylinder with a piston and containing a gas. The refrigerator extracts heat from a cold bath at temperature $T_1$ when we expand the gas by pulling out the piston. The cylinder is then disconnected from the cold bath and isolated thermally. The gas is then compressed until achieves temperature $T_2$ equal to that of the room and it is then put in thermal contact with the room and it is further compressed slowly, releasing heat into the room. The movement of the piston during the entire process has to be slow, as not to produce pressure and heat waves inside the gas, but this is not enough. We must also ensure that the point at which the gas is compressed while the cylinder is thermally isolated is precisely engineered to correspond to the gas reaching room temperature.

\section{Carnot efficiency for the refrigerator}

The upshot of the above arguments is that as far as the question of efficiency in the reversible regime is concerned, most of the details of the refrigerator are irrelevant and it all reduces to three simple equations: Equation (\ref{energies}) that describes the basic built of the fridge, equation (\ref{ratios}) that describes the connection between the heat flows and the basic construction of the refrigerator and equation (\ref{working_regime3_fridge}) that describes the reversible regime.

From (\ref{energies}) and (\ref{working_regime3_fridge}) we obtain

\beq{{E_1}\over{T_1}}+{{E_3}\over{T_3}}={{E_1+E_3}\over{T_2}}\eeq
which then, using (\ref{ratios}) leads to
\beq{{Q_1}\over{T_1}}+{{Q_3}\over{T_3}}={{Q_1+Q_3}\over{T_2}}\eeq
which is the relation that connects $Q_3$, the heat extracted by the fridge from the cold bath and $Q_1$ the heat extracted from the hot bath that drives the refrigerator. This expression is identical to that of a classical reversible refrigerator that works between these temperatures (see \cite{lps}).

\section{A quantum self-contained heat engine}

In \cite{heat_engine} a model for a quantum self-contained heat engine was proposed. The engine consists by two qubits, 1 and 2, in contact with a hot bath, $T_1$ and a cold bath, $T_2<T_1$. The energy separation between the ground state $|0\ra_i$ and the excited state $|1\ra_i$ of qubit $i$ is $E_i$. The engine lifts a weight in equal steps; the energy difference between two subsequent positions, $|n\ra_w$ and $|n+1\ra_w$ is $E_3$. The weight is not connected to any heat bath.

The engine is constructed such that

\beq E_1=E_2+E_3\eeq.

Due to the above constraint, the states $|10n\ra$ and $|01n+1\ra$ are degenerate where $|10n\ra$ stands for $|1\ra_1|0\ra_2|n\ra_w$ and so on. Again, an interaction Hamiltonian is added, of magnitude smaller than that of the free hamiltonians. Due to this, it can only induce transitions between the degenerate eigenstates.

Again, the basic idea of the engine is to make the transition  $|10n\ra\rightarrow|01n+1\ra$ in which the weight is lifted more favorable than the reverse transition. The condition for this is

\beq e^{-{{E_1}\over{kT_1}}} > e^{-{{E_2}\over{kT_2}}}\label{working_regime}\eeq
which leads to
\beq{{E_1}\over{T_1}}<{{E_2}\over{T_2}}\label{working_regime2}.\eeq
As in the case of the refrigerator, the reversibility condition is when the froward and backward transitions are equally probable, i.e.
\beq{{E_1}\over{T_1}}={{E_2}\over{T_2}}\label{working_regime2}.\eeq

Finally, the interaction imposes the constraint that whenever the wight is lifted and it gains energy $E_3$, qubit 2 gains energy $E_2$ and qubit 1 loses energy $E_1$ and vice versa. While we cannot tell anything about the time scales involved without more information about the parameters of the device, it is clear that the ratio between the heat  $Q_1$ extracted by qubit 1 from the hot bath, the heat $Q_2$ dumped by qubit 2 into the cold bath and the work gained by the weight are in the same ratio as the energies gained and lost in one transition:

\beq Q_1 : Q_2 : W = E_1 : E_2 : E_3\eeq

Putting all this together we obtain

\beq{{Q_1}\over{T_1}}={{Q_1-W}\over{T_2}}\eeq
which can be arranged to read
\beq {W\over{Q_1}}=1-{{T_2}\over{T_1}}\eeq
the well known expression for the Carnot efficiency of a simple classical heat engine.

\section{Conclusion}

To conclude, we presented the basic principles for the functioning of quantum self-contained heat machines. While to find out the exact working parameters outside of the reversible regime is complicated and depends on all the parameters of the problem, all these details become irrelevant in the reversible regime. The three simple and very intuitive equations are enough to tell that the quantum engines reach ideal Carnot efficiency

\section {Further considerations}

Our general considerations above also allow us to immediately tell a couple of things about the working point far from reversibility.

Suppose that the thermalisation model is the one in \cite{lps} in which with probability $p_i$ per unit time the state of the qubit gets replaced by the thermal state $\tau_i$. Furthermore suppose that the steady state is also diagonal in the same basis, $|0\ra_i$, $|1\ra_i$. The heat extracted per unit time is $p_i$ times the energy gained/lost when a replacement is made. This energy is equal to the difference of the mean energies in the steady state $\rho^S_i$ and $\tau_i$;  which is equal to $E_i$ times $\delta q_i$ the change in probability for the excited state. Hence
\beq Q_i=p_iE_i\delta q_i\eeq
But taking into account that the ratio of heat flows is the same as the ratio of energies (\ref{ratios}) we obtain that
\beq p_1\delta q_1=p_2\delta q_2=p_3\delta q_3\eeq
which is confirmed by the explicit form deduced in \cite{lps}.

\end{document}